\documentclass{appolb}
\usepackage{graphicx}
\usepackage{subfigure}

\preprint{}

\begin{document}

\title{{\bf Soft-hadronic observables for relativistic heavy-ion collisions at RHIC and LHC}\thanks{Talk presented  by WF at the 
{\em Cracow Epiphany Conference on LHC Physics}, Krak\'ow, Poland, January 4 -- 6, 2008}
\thanks{Partly supported by the Polish Ministry of Science and Higher Education, grants N202 153 32/4247 and N202 034 32/0918, and by the U.S. National Science Foundation under Grant No. PHY-0653432.}}
\author{Wojciech Florkowski $^{1,2}$, Mikolaj Chojnacki $^1$, \\ Wojciech Broniowski $^{1,2}$ and Adam Kisiel $^{3,4}$ 
\address{
$^1$ H. Niewodnicza\'nski Institute of Nuclear Physics, Polish Academy of Sciences, PL-31342 Krak\'ow, Poland \\
$^2$ Institute of Physics, Jan Kochanowski University, PL-25406~Kielce,~Poland \\
$^3$ Faculty of Physics, Warsaw University of Technology, PL-00661 Warsaw, Poland \\
$^4$ Department of Physics, Ohio State University, 1040 Physics Research Building, 191 West Woodruff Ave., Columbus, OH 43210, USA}}

\maketitle

\begin{abstract}
The relativistic hydrodynamics together with the single-freeze-out statistical hadronization model is used to describe the midrapidity hadron production in ultra-relativistic heavy-ion collisions at RHIC and LHC. At the highest RHIC energy our approach gives a quite satisfactory global description of soft hadronic observables including the HBT radii. With the increased initial energy, from RHIC to LHC, we expect the saturation of the pion elliptic flow and the moderate increase of the correlation radii. 
\end{abstract}


\thispagestyle{empty}

\section{Introduction}
\label{sect:intro}

The basic features of the soft hadron production, such as the transverse-momentum spectra and the elliptic flow coefficient $v_2$, are successfully described by the hydrodynamic models \cite{Huovinen:2001cy,Teaney:2000cw,Teaney:2001av,Hirano:2002ds,Kolb:2002ve,Hama:2005dz,Eskola:2005ue,Nonaka:2006yn}, for recent reviews see \cite{Kolb:2003dz,Huovinen:2003fa,Shuryak:2004cy,Muller:2007rs,Nonaka:2007nn}. In this paper, encouraged by the success of this framework, we use the newly developed hydrodynamic approach \cite{Chojnacki:2004ec,Chojnacki:2006tv,Chojnacki:2007jc} linked to the statistical-hadronization Monte-Carlo model {\tt THERMINATOR} \cite{Kisiel:2005hn} to globally describe the soft hadron production at the highest RHIC energies and  to make predictions for the ultra-relativistic heavy-ion experiments at the Large Hadron Collider (LHC). 

The distinctive features of our approach are the following: 1) The realistic equation of state is used which interpolates smoothly between the lattice results \cite{Aoki:2005vt} and the hadron gas model. In this way a cross-over phase transition is included in the dynamics of the system.  2) The statistical hadronization Monte-Carlo model \cite{Kisiel:2005hn} is applied to describe hadron production and resonance decays. The use of the full set of hadronic resonances allows to incorporate a relatively high freeze-out temperature, $T_f \sim$ 150 MeV, which helps to describe the magnitude of the HBT radii.  3) The two-particle method including Coulomb effects is used to calculate the correlation functions. This method mimics, as closely as possible, the procedures used in the experimental measurements.

The drawback of our approach is that we do not include the elastic rescattering in the final state (after the chemical freeze-out). As may be concluded from the results presented in Ref.~\cite{Nonaka:2006yn} such effects are small for pions. On the other hand, they are more important for protons and may be responsible for differences betweeen the model results and the data in this case.

Studying the hadron production at the highest RHIC energies, we make an attempt to describe consistently not only the transverse-momentum spectra and the elliptic flow coefficient $v_2$ but also the HBT correlation radii. We find that a global fit to pion and kaon observables is possible at the level of 10-15\%, which we find quite satisfactory, having in mind systematic uncertainties in various elements of the approach.

By means of simple extrapolations to higher energies (i.e., increasing the total inelastic nucleon-nucleon  cross section, initial temperature, and the contributions from hard scattering to the initial entropy) we make predictions for soft hadronic observables at LHC. We note that the similar predictions of other hydrodynamic models for phenomena expected at the LHC have been recently presented and summarized in Ref.~\cite{Abreu:2007kv}.

\section{Initial conditions}

The hydrodynamic evolution starts at the proper time $\tau_0$ = 1 fm. We assume that there is no initial transverse flow formed at this moment. The initial entropy density profile in the transverse plane is proportional to the linear combination of the wounded-nucleon density \mbox{$\rho_{\rm W}\left( \vec{x}_{T } \right)$} and the density of binary collisions \mbox{$\rho_{\rm bin}\left( \vec{x}_{T } \right)$}, namely
\begin{equation}
  s \left( \vec{x}_{T } \right)   \propto 
  \rho \left( \vec{x}_{T } \right) = 
  \frac{1-\kappa}{2} \, \rho_{\rm W}\left( \vec{x}_{T } \right) + \kappa \rho_{\rm bin} \left( \vec{x}_{T } \right). 
  \label{initialeps2}
\end{equation}
The case $\kappa=0$ corresponds to the standard wounded-nucleon model \cite{Bialas:1976ed}, while $\kappa=1$ would include  the binary collisions only. The PHOBOS analysis \cite{Back:2001xy,Back:2004dy} of the particle multiplicities yields $\kappa=0.12$ at $\sqrt{s_{NN}}=17~{\rm GeV}$ and $\kappa=0.14$ at $\sqrt{s_{NN}}=200~{\rm GeV}$. In this paper we assume $ \kappa=0.14$ for RHIC and $ \kappa=0.2$ for LHC. 

The wounded-nucleon and the binary-collisions densities in Eq.~(\ref{initialeps2}) are obtained from the optical limit of the Glauber model, which is a very good approximation for not too peripheral collisions \cite{Miller:2007ri}. In the calculations of those densities we use two values of the nucleon-nucleon total inelastic cross section: $\sigma = 42~{\rm mb}$ for RHIC and $\sigma = 63~{\rm mb}$ for LHC. Different values of the nucleon-nucleon cross section imply different relations between the centrality classes and the impact parameters. In this paper we restrict our presentation to the discussion of the centrality class $c$ = 20-30\% and use the values: \mbox{$b$ = 7.16 fm} for RHIC and $b$ = 7.60 fm for LHC. 
We note that different nuclear density profiles are used for RHIC (gold on gold collisions) and LHC (lead on lead collisions).

The overall normalization of the entropy density is determined by the value of the initial central temperature $T_i$. This parameter controls mainly the absolute yields of the particles. From the fits to the RHIC data we find $T_i$ = 320 MeV. We expect that a higher initial temperature will be achieved at LHC, hence we use $T_i$ = 450 MeV in this case. The other possible values of the initial temperature at LHC, as well as other centralities,  were analyzed thoroughly in Ref. \cite{Chojnacki:2007rq}. 

The input parameters for the RHIC and LHC calculations are summarized in Table~\ref{tab:rhiclhc}. We note that the initial conditions for hydrodynamics used in this paper may be regarded as the standard approach. Very recently, we have found that the departure from such a standard form, i.e., the use of the Gaussian energy profiles with width parameters obtained from {\tt GLISSANDO} \cite{Broniowski:2007nz}, leads to even better description of the RHIC data 
\cite{Broniowski:2008vp} (see also Ref.~\cite{Gyulassy:2007zz} in this context). The consequences of the modified initial conditions will be discussed in the forthcoming publication, while in the present paper we restrict our consideration to the typical initial conditions. 

\begin{table}[b]
\begin{center}
\caption{Comparison of the RHIC and LHC initial physical parameters used in the present calculation. 
\label{tab:rhiclhc}}
\begin{tabular}{|c|c|c|c|c|c|}
\hline
     & $\sigma$ & $\kappa$ & $T_i$ &  $b$ (for $c$ = 20 - 30\%) \\
\hline
RHIC & 42 mb    & 0.14     & 320 MeV   & 7.16 fm \\
LHC  & 63 mb    & 0.20     & 450 MeV   & 7.60 fm \\
\hline
\end{tabular}
\end{center}
\end{table}

\section{Hydrodynamic evolution}

Our hydrodynamical code describes boost-invariant systems with zero baryon chemical potential. However, its advantage is that it incorporates the state-of-the-art equation of state \cite{Chojnacki:2007jc}. At high temperatures the thermodynamic variables agree with the recent lattice results \cite{Aoki:2005vt} while at low temperatures they agree with the resonance gas model. In the cross-over region near $T_c \sim$ 170 MeV, a simple interpolation between the high- and low-temperature equations of state is constructed. In accordance with the present understanding, no phase transition but rather a smooth cross-over takes place in the vicinity of the critical temperature. 

Since we concentrate on the midrapidity region, $|y|<1$, the boost invariance seems to be a reasonable approximation for the highest RHIC and LHC energies \cite{Bearden:2003fw,Bearden:2004yx}. Recently, the interesting results were obtained in Ref. \cite{Bozek:2007qt} (see also these Proceedings \cite{Bozek:2008bh}), where the impact of  the shear viscosity on the longitudinal (not boost-invariant) dynamics was analyzed. In the scenario where the effects of the longitudinal acceleration are compensated by the shear viscosity, the boost-invariance may be a quite good approximation for the region $|y|<1$ even if the system is not globally boost-invariant.

\begin{figure}[t]
\begin{center}
\subfigure{\includegraphics[angle=0,width=0.45\textwidth]{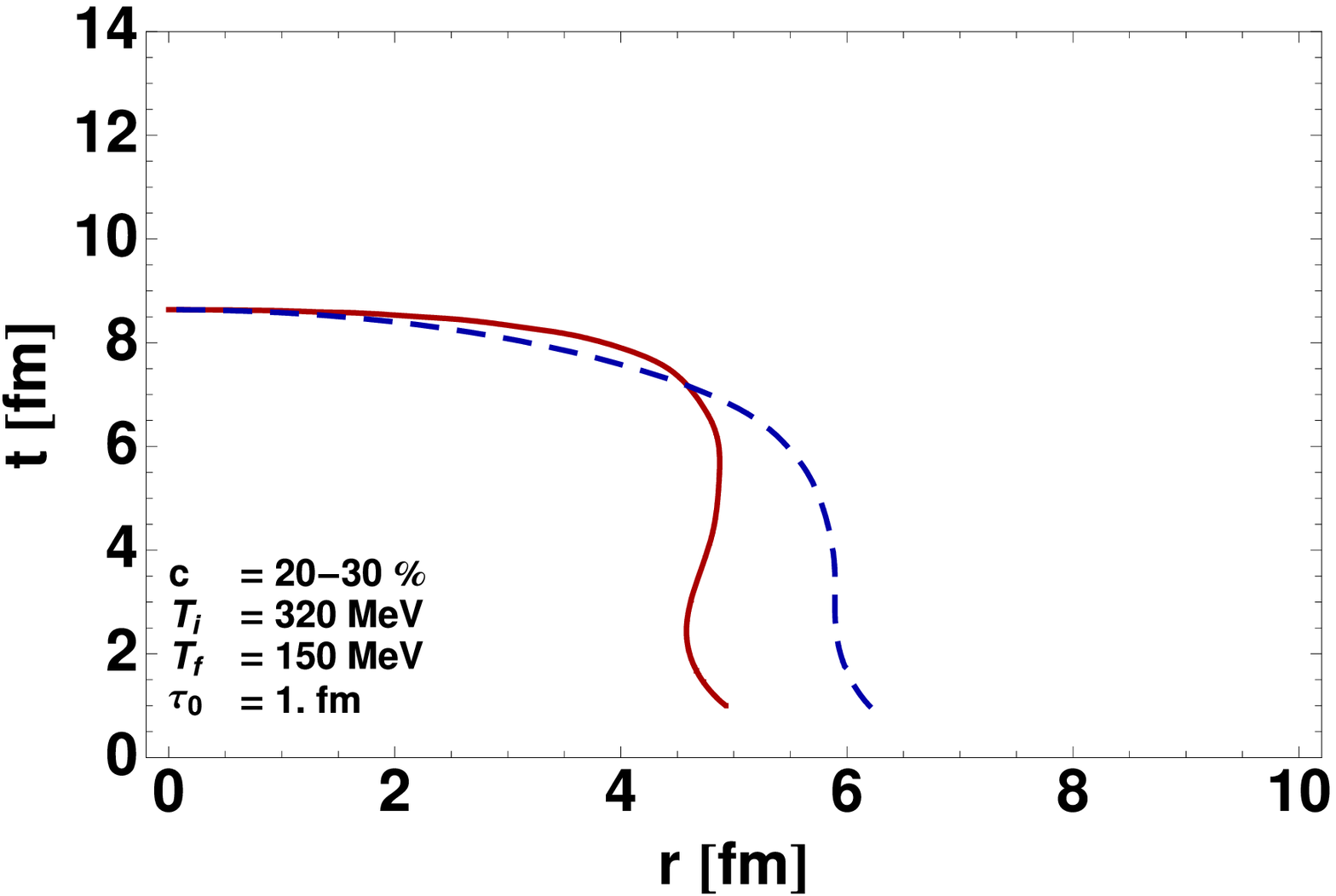}} 
\subfigure{\includegraphics[angle=0,width=0.45\textwidth]{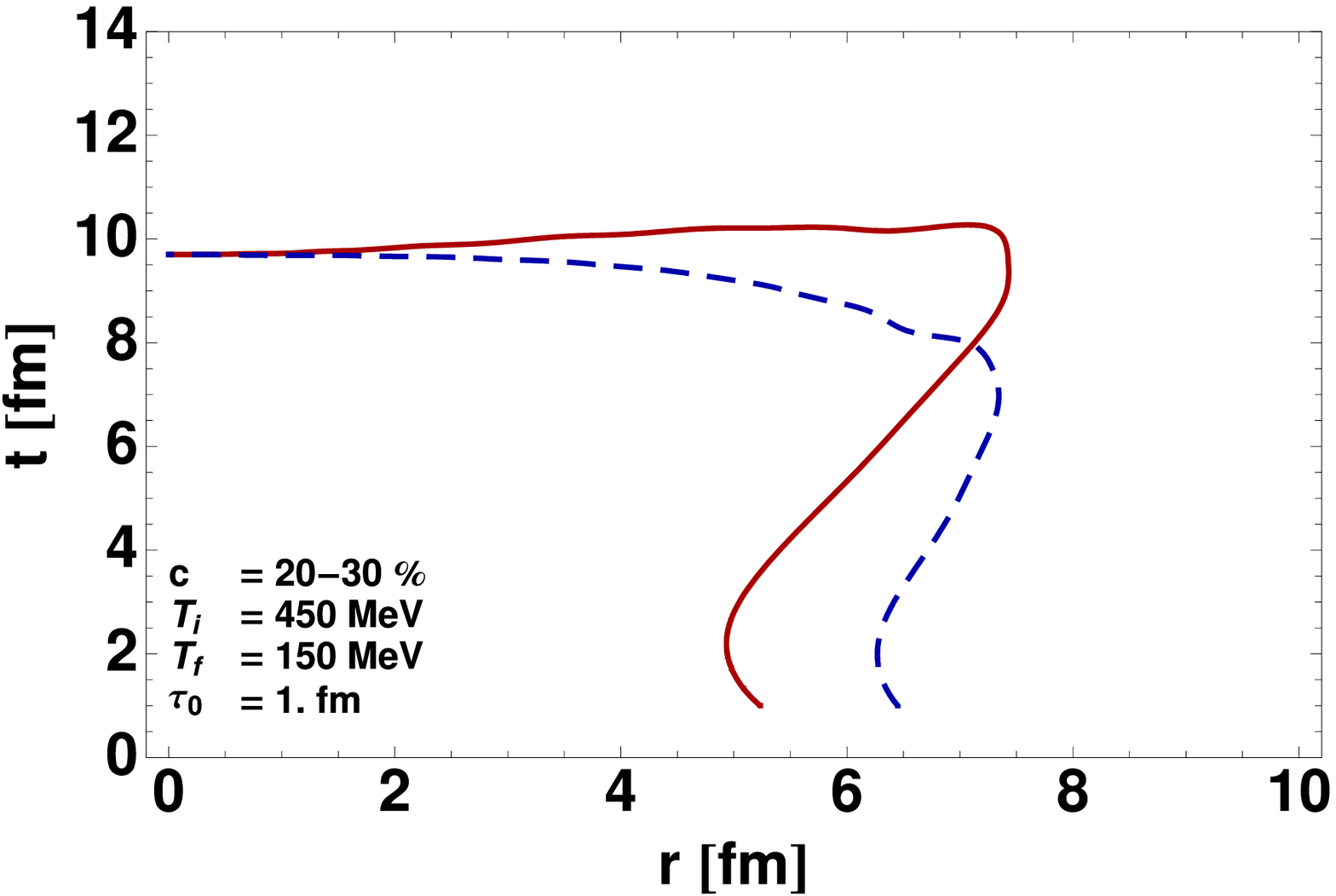}} \\
\end{center}
\caption{{\bf Left:} The freeze-out curves for RHIC: \mbox{$T_i = 320~{\rm MeV}$}, \mbox{$T_f = 150~{\rm MeV}$}, and \mbox{$b = 7.16~{\rm fm}$}. 
{\bf Right:} The freeze-out curves for LHC:  \mbox{$b = 7.6~{\rm fm}$},  \mbox{$T_i = 450~{\rm MeV}$}, and \mbox{$T_f = 150~{\rm MeV}$}. In both figures the solid lines describe the in-plane profiles, while the dashed lines describe the out-of-plane profiles.  }
\label{fig:hs}
\end{figure}

The hydrodynamic evolution proceeds until the temperature drops down to the freeze-out temperature $T_f$. The two temperatures, $T_i$ and $T_f$, are the basic fit parameters of the approach. The hydrodynamic equations are solved with the technique described in detail in Refs.~\cite{Chojnacki:2004ec,Chojnacki:2006tv,Chojnacki:2007jc} which is a generalization of the method introduced by Baym et al. in Ref. \cite{Baym:1983sr}. Entropy conservation is used as a numerical test of the numerical code. It is satisfied at the relative level of $10^{-4}$ or better.

\begin{table}[b]
\begin{center}
\caption{Comparison of the RHIC and LHC freeze-out thermodynamic parameters used in the present calculation. 
\label{tab:rhiclhc2}}
\begin{tabular}{|c|c|c|c|c|}
\hline
     & $T_f$ & $\mu_B$ & $\mu_S$ &   $\mu_{I_3}$  \\
\hline
RHIC & 150 MeV   & 28.5 MeV   & 9 MeV    & 0.9 MeV \\
LHC  & 150 MeV    & 0.8 MeV   & 0   & 0\\
\hline
\end{tabular}
\end{center}
\end{table}

\section{Results for RHIC and LHC}

The freeze-out hypersurface and the flow profile obtained from the hydrodynamic evolution are used as input for the thermal event generator {\tt THERMINATOR} \cite{Kisiel:2005hn}. The examples of the freeze-out hypersurfaces obtained for the centrality class $c$ = 20-30\%  are shown in Fig.~\ref{fig:hs}. The higher initial temperature expected at LHC leads evidently to larger sizes of the system and larger transverse flow (the last effect is indicated by the extended shape of the freeze-out hypersurface). 

Below we present our results describing transverse-momentum spectra, the elliptic flow, and the HBT radii, which were obtained with the two freeze-out hypersurfaces shown in Fig.~\ref{fig:hs}. The {\tt THERMINATOR} implements the statistical hadronization, accounting for a complete treatment of hadronic resonances (the included resonances and their branching ratios are the same  as in the {\tt SHARE} package \cite{Torrieri:2004zz}). Rescattering after the chemical freeze-out is not incorporated, which is a reasonable approximation for pions. Basic physical observables are calculated from the sample of events generated by {\tt THERMINATOR}. 
According to the results of the statistical models \cite{Florkowski:2001fp,Braun-Munzinger:2001ip,Baran:2003nm,Andronic:2005yp}, we assume the specific values of the chemical potentials at freeze-out, see Table~\ref{tab:rhiclhc2}. Since these values are small, as compared to the temperature which is above 150 MeV, the effects of the chemical potentials may be neglected in the hydrodynamic evolution. 

\begin{figure}[t]
\begin{center}
\subfigure{\includegraphics[angle=0,width=0.48\textwidth]{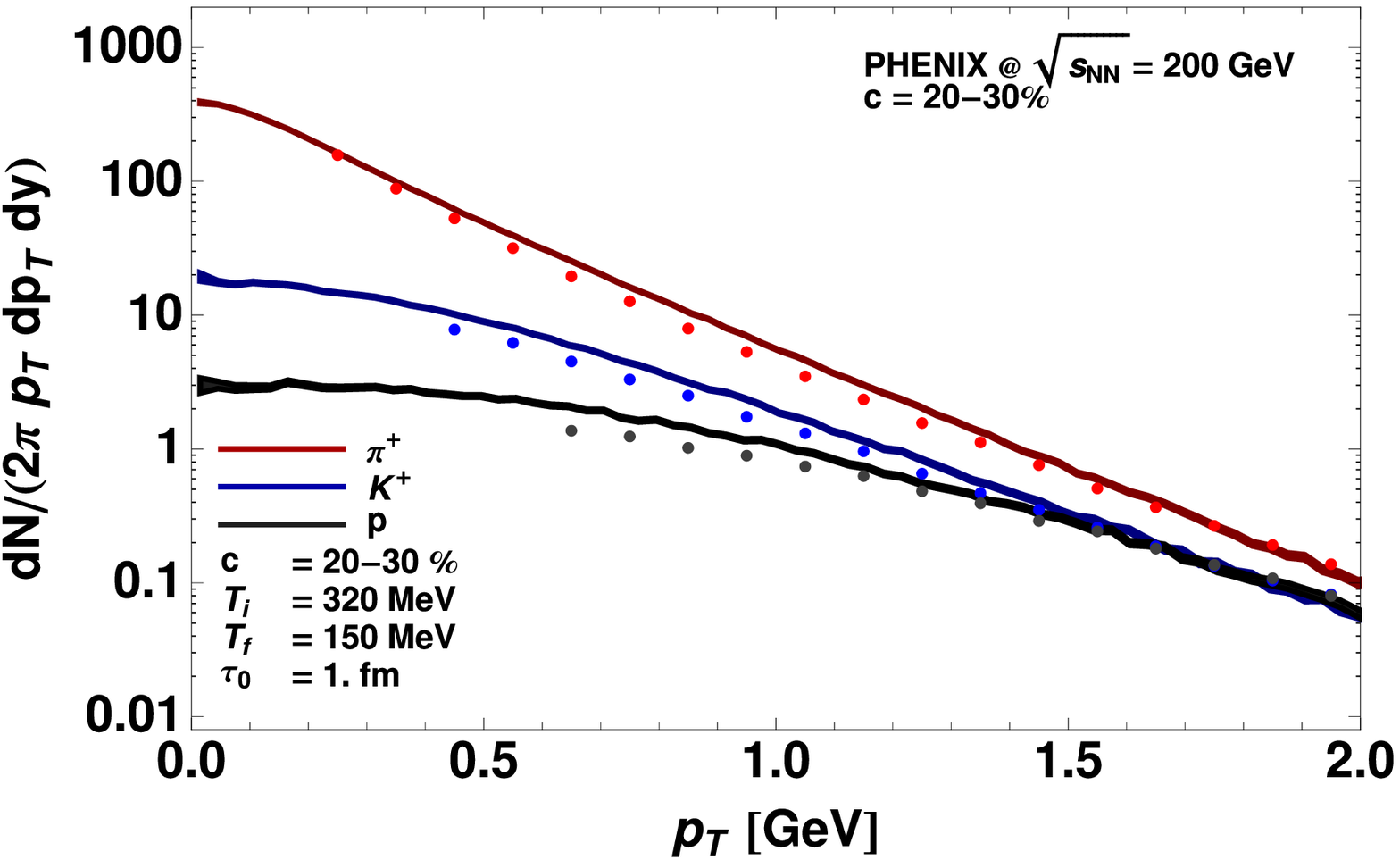}} 
\subfigure{\includegraphics[angle=0,width=0.48\textwidth]{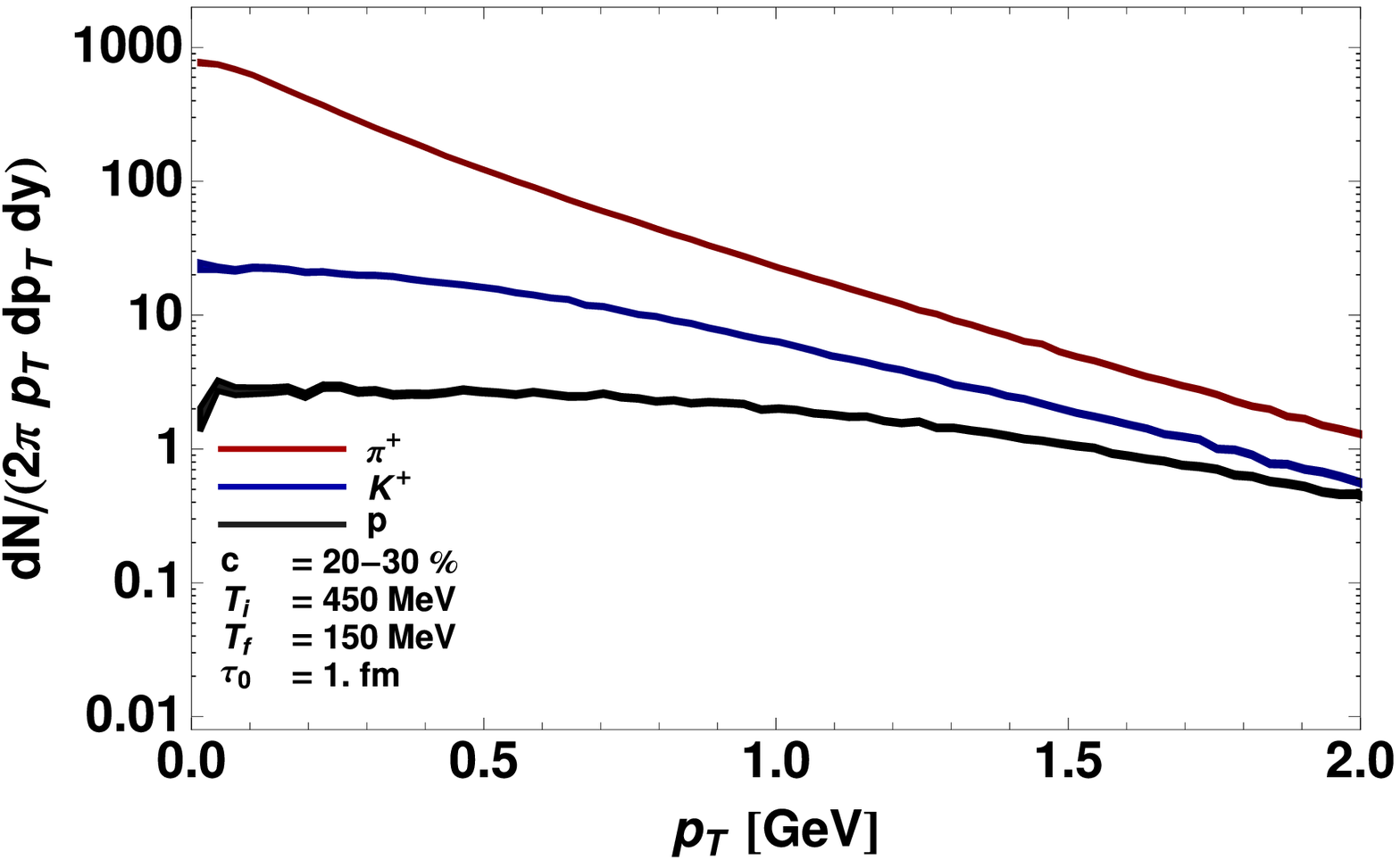}} \\
\end{center}
\caption{Transverse-momentum spectra of $\pi^+$, $K^+$, and protons. 
{\bf Left:} The PHENIX experimental results \cite{Adler:2003cb} (points) are compared to the RHIC model calculations (solid lines). {\bf Right:} The corresponding model calculations for the LHC case. The values of the model parameters are the same as in Fig. \ref{fig:hs}.}
\label{fig:ptspec}
\end{figure}

\begin{figure}[t]
\begin{center}
\subfigure{\includegraphics[angle=0,width=0.45\textwidth]{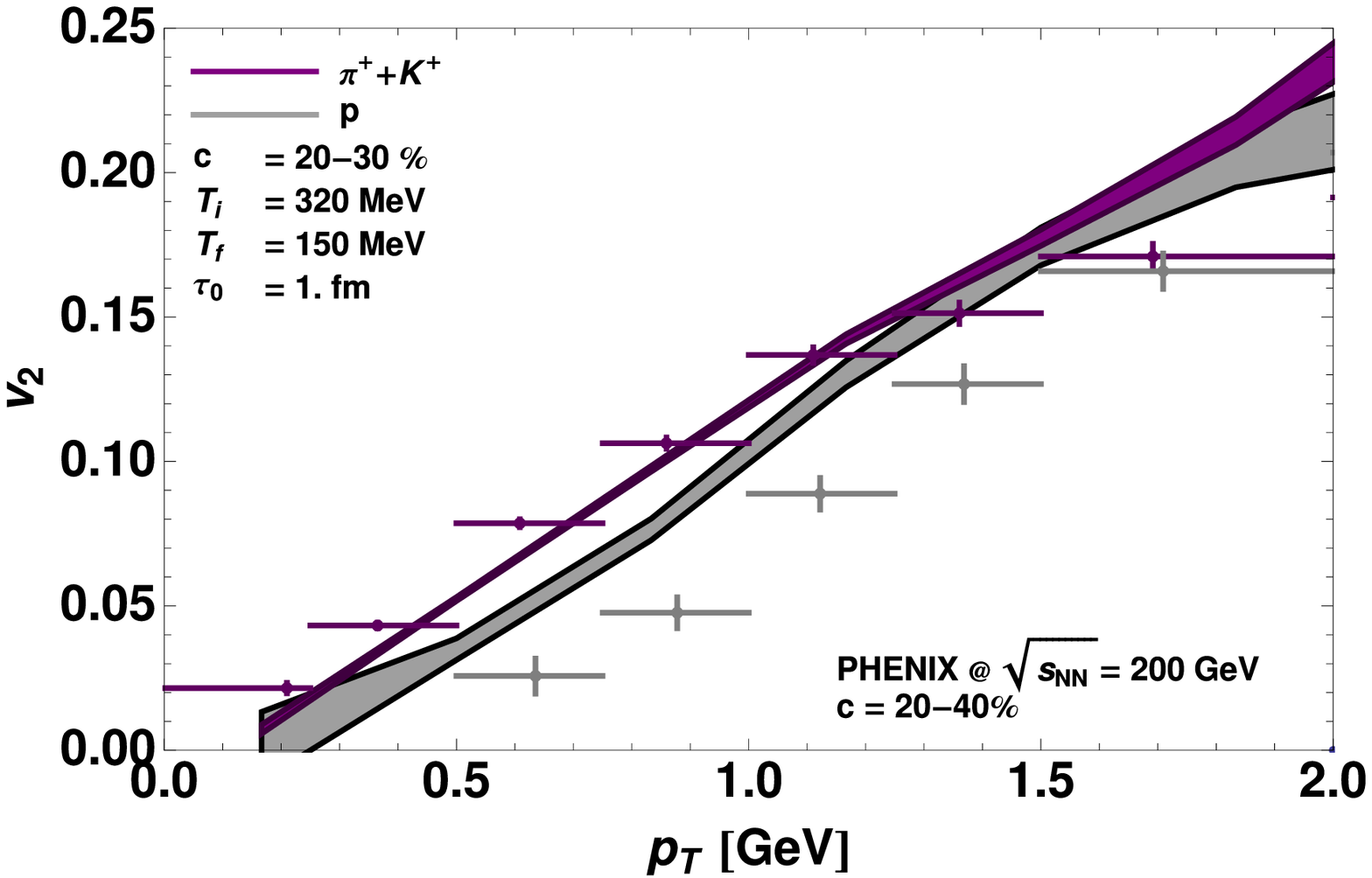}} 
\subfigure{\includegraphics[angle=0,width=0.45\textwidth]{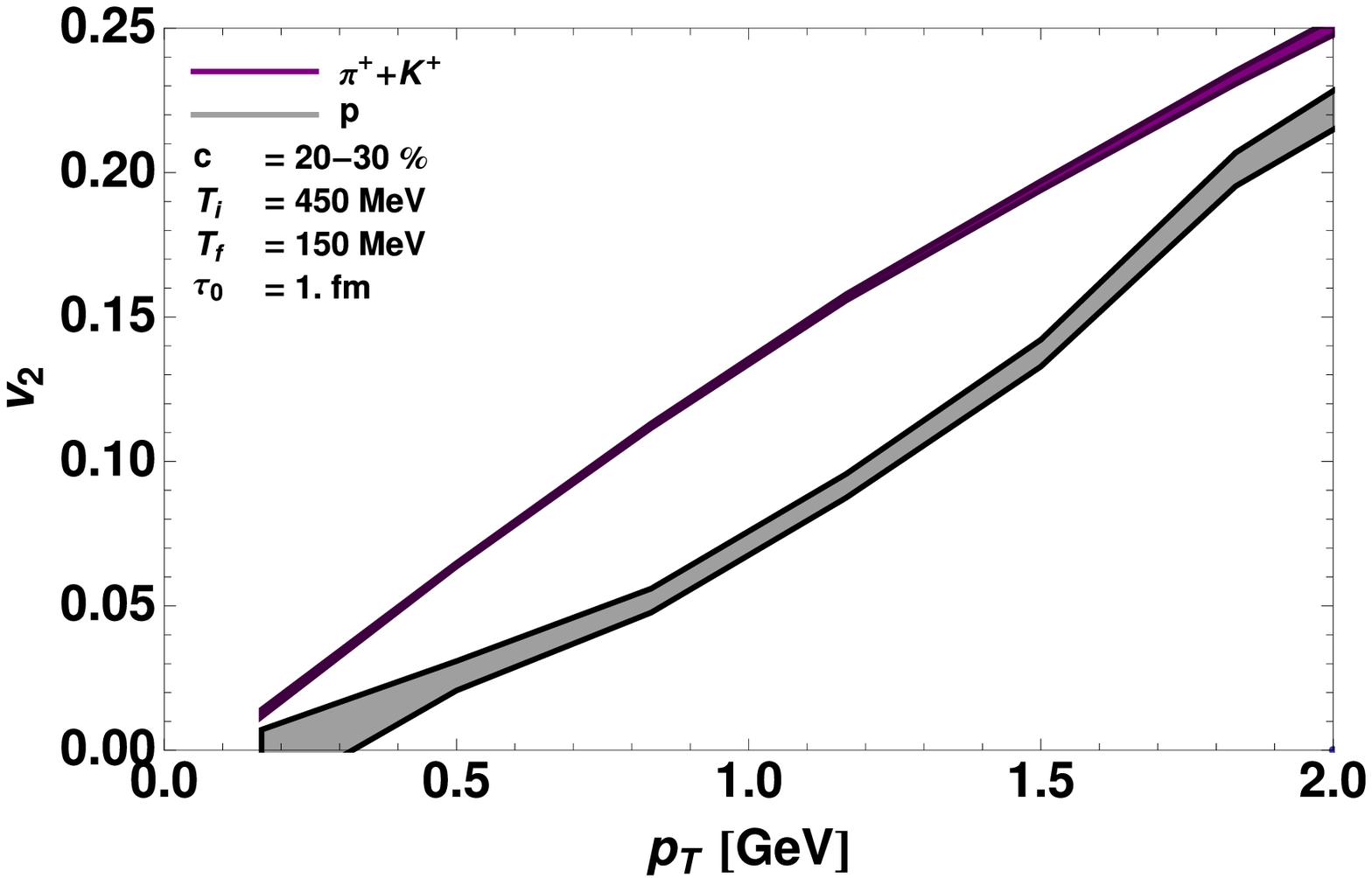}} \\
\end{center}
\caption{The elliptic flow coefficient $v_2$.
{\bf Left:} The PHENIX experimental results \cite{Adler:2003kt} (points with horizontal bars) are compared to the RHIC model calculations (solid lines). {\bf Right:} The corresponding model calculations for the LHC case.  The values of the model parameters are the same as in Fig. \ref{fig:hs}.}
\label{fig:v2}
\end{figure}


In Fig.~\ref{fig:ptspec} we show our results describing the transverse-momentum spectra of pions, kaons, and protons. The proton spectra are corrected for the weak $\Lambda$ decays. 
In the left part of Fig.~\ref{fig:ptspec} one can see that the pion, kaon, and proton transverse-momentum spectra measured for Au+Au collisions at $\sqrt{s_{NN}}= 200~{\rm GeV}$ and the centrality class 20-30\% by PHENIX (points) are reasonably  well reproduced by the model calculations. In the right part of Fig.~\ref{fig:ptspec} we present the corresponding results for LHC. One observes higher multiplicities of the produced particles (the effect caused by a higher initial temperature) and flatter spectra (the effect caused by the larger transverse flow).

\begin{figure}[t]
\begin{center}
\subfigure{\includegraphics[angle=0,width=0.45\textwidth]{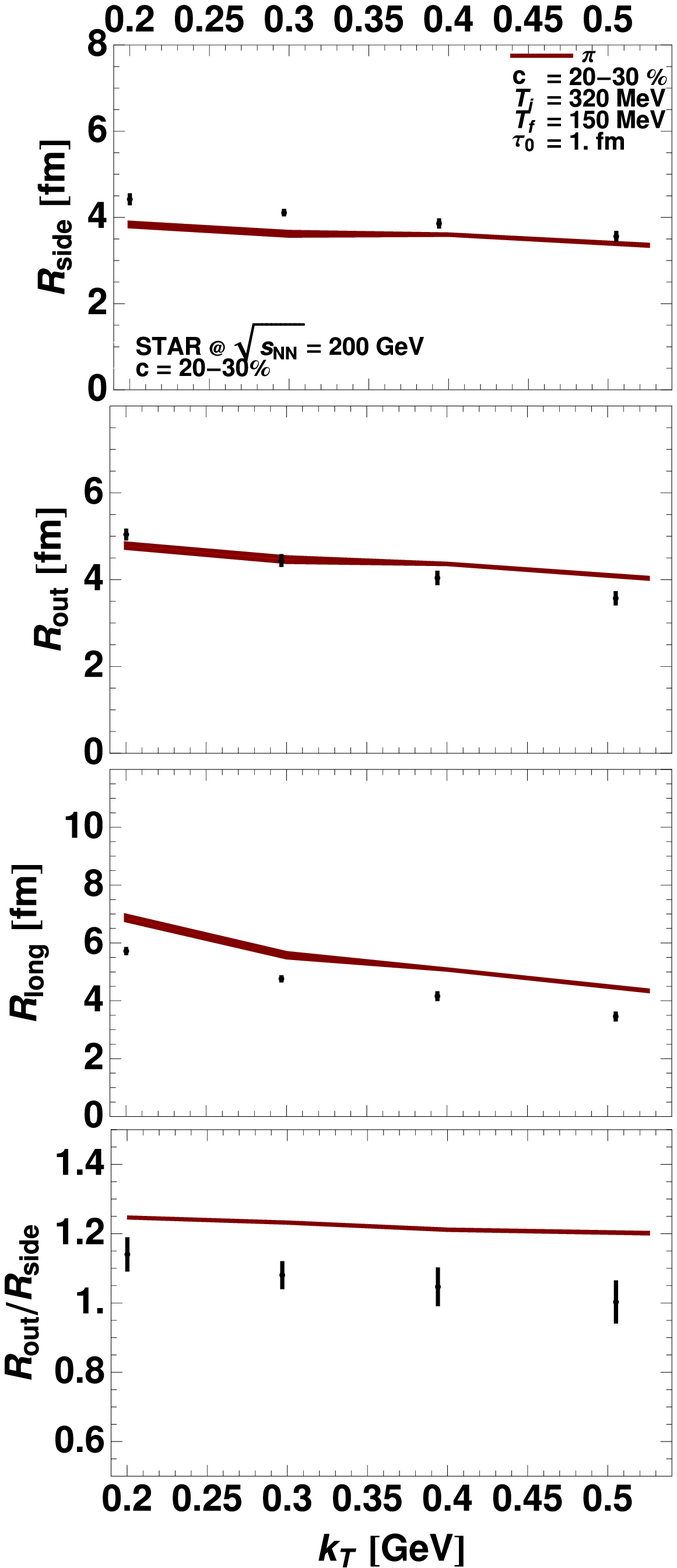}} 
\subfigure{\includegraphics[angle=0,width=0.45\textwidth]{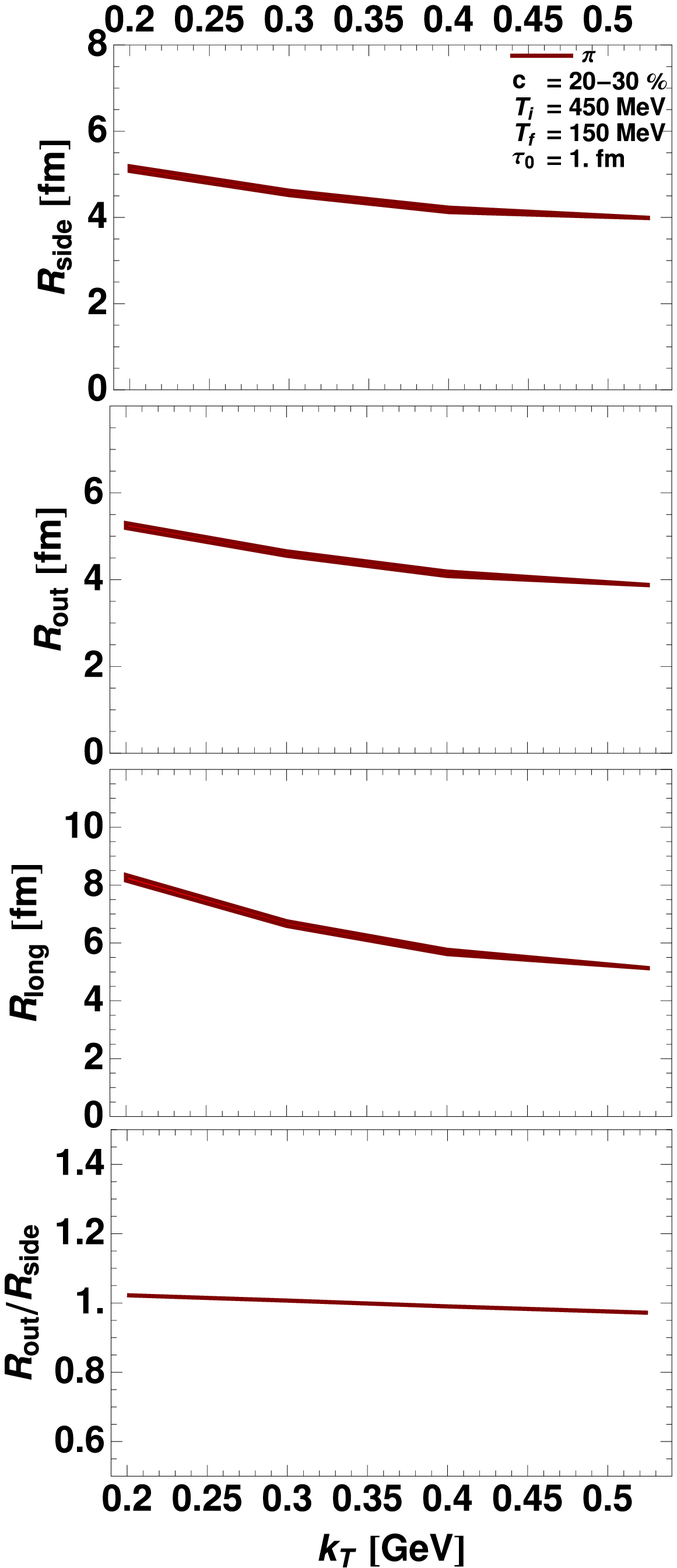}} \\
\end{center}
\caption{The pion HBT radii plotted as the function of the transverse-momentum of pion pairs.
{\bf Left:} The STAR experimental results \cite{Adams:2004yc} (points) compared to the RHIC model calculations (solid lines). {\bf Right:} The corresponding model calculations for the LHC case.  The values of the model parameters are the same as in Fig. \ref{fig:hs}.}
\label{fig:hbt}
\end{figure}


In Fig.~\ref{fig:v2} we present our results describing the elliptic flow coefficient $v_2$. In the left part of Fig.~\ref{fig:v2} the values measured by PHENIX \cite{Adler:2003kt} at $\sqrt{s_{NN}}= 200~{\rm GeV}$ and the centrality class 20-40\% are indicated by the upper (pions + kaons) and lower (protons) points, with the horizontal bars indicating the $p_T$ bin. The corresponding model calculations are indicated by the solid lines, with the bands displaying the statistical error of the Monte-Carlo method.  We observe that the $v_2$ of pions+kaons agrees with the data. On the other hand, the model predictions for $v_2$ of protons is too large. The discrepancy is probably caused by the final-state elastic interactions, not included in our approach. In the right part of Fig.~\ref{fig:v2} we show our predictions for LHC. They indicate the saturation of the elliptic flow of light particles for a given initial space asymmetry. On the other hand, the proton elliptic flow is significantly reduced. This observation is consistent with the findings of Kestin and Heinz discussed 
in Ref.~\cite{Abreu:2007kv}.


Our results describing the HBT radii are shown in Fig.~\ref{fig:hbt}. The femtoscopic observables are evaluated  with the help of the two-particle method accounting for the effects of resonance decays and the Coulomb final-state interactions. From the left part of Fig.~\ref{fig:hbt} we conclude that the pionic HBT radii measured by STAR at centrality 20-30\% \cite{Adams:2004yc} are reasonably reproduced in our approach. The ratio $R_{\rm out}/R_{\rm side}$ is about 1.2-1.25, which is still significantly above the experimental ratio, but considerably better than in many other hydrodynamic approaches. We note, that the use of the Gaussian initial condition \cite{Broniowski:2008vp} allows for a much better description of the HBT radii, including the ratio of $R_{\rm out}/R_{\rm side}$, as well as the azimuthally sensitive femptoscopy. In the right part of Fig.~\ref{fig:hbt} we present the corresponding results for LHC. We observe the moderate increase of all the radii with the ratio $R_{\rm out}/R_{\rm side}$ very close to one.

\section{Conclusions}

We summarize our findings with the two observations: 

\begin{itemize}

\item{} The standard relativistic hydrodynamics followed by the statistical hadronization assuming single-freeze-out scenario \cite{Florkowski:2001fp,Broniowski:2001we,Broniowski:2002nf} describes well the global features of soft hadron production in heavy-ion collisions at the highest RHIC energies \cite{Chojnacki:2007rq}. This picture may be still improved by imposing different initial conditions, as argued in Ref. \cite{Broniowski:2008vp}.

\item{} The increase of the initial temperature, which is the main expected effect in the transition from RHIC to LHC,  yields a rather smooth change of the basic soft-physics observables. The observed multiplicities, spectra, and HBT radii reflect the increased values of entropy and collective flow \cite{Chojnacki:2007rq}. 

\end{itemize}

Certainly,  the LHC results coming in the next years will verify this picture. We should remember that the RHIC measurements did not meet earlier expectations. Similar situation may happen again in the case of LHC leaving us with the new challenges.


\end{document}